\documentstyle[12pt]{article}

\begin{document}

\begin{titlepage}
\begin{center}
        {\large \bf PARTICLE CREATION IN A UNIVERSE FILLED BY \\
          RADIATION AND DUST-LIKE MATTER}

\vspace{1.0 cm}

V\'{\i}ctor M. Villalba\\
Centro de F\'{\i}sica,\\
            Instituto Venezolano de Investigaciones Cient\'{\i}ficas
(IVIC)\\
            Apartado 21827, Caracas $1020A$, Venezuela

\end{center}
\vspace{1.5 cm}

\begin{abstract}
In this article the particle creation process of scalar and spin 1/2
particles in a spatially open cosmological model associated with a
universe
filled by radiation and dustlike matter. The Klein-Gordon and Dirac
equations
are solved via separation of variables. After comparing  the{\it \ in
}and%
{\it \ out} vacua, we obtain that the number of created particles
corresponds to
 Planckian and Fermi-Dirac
distributions  for the scalar and Dirac cases
respectively.
\end{abstract}
\vspace{1.5cm}

\end{titlepage}

\section{Introduction}

\setcounter{equation}{0}

During the last two decades there has been an increasing interest in the
study of quantum effects in cosmological backgrounds. Since the
appearance
of the pioneer works by Parker \cite{Parker1} and Grib {\it et al} \cite
{Grib} among others, a great body of papers have been published on the
problem of scalar and spin 1/2 particle creation in different
cosmological
configurations. More recently there have been some attempts to analyze
in
this framework the problem of particle creation in the vicinity of
cosmic
strings \cite{Parker2,Sahni}

Since the problem of particle creation in curved space-times is closely
related to the definition of particle in a curved background, different
approaches and interpretations have appeared during these last two
decades,
being the most popular the adiabatic and the corpuscle interpretations,
an
overview of the most important results and approaches can be found in
\cite
{Birrel} and \cite{Grib2}. Despite the great effort devoted to a better
understanding of quantum phenomena in expanding cosmological universes,
the
number of articles discussing concrete models is relatively scarce. This
is
due to the fact that, in most of the cases, the relativistic wave
equations
in curved backgrounds are very difficult to solve and therefore is
necessary
to reduce the original problem to a simpler one.

Recently, the exact solution of Klein-Gordon equation in an expanding
(asymptotically flat) cosmological universe filled with radiation has
been
reported \cite{Costa}, the authors obtain a thermal planckian
distribution
of particles created when they compare the ''in'' and ''out'' vacua via
the
Bogoliubov coefficients. The model presented in \cite{Costa} encourages
us
to analyze more complex configurations.

A good scenario for discussing the problem of scalar and spin 1/2
particle
creation in a cosmological background, is an isotropic universe with
zero
cosmological constant $(\Lambda ~=~0)$ filled with a mixture of
radiation
and dust-like matter \cite{Chernin,Kharbediya} in a spatially open
Friedmann
Robertson-Walker model of the form
\begin{equation}
\label{Fried}ds^2~=-dt^2+a^2(t)\left( d\chi ^2+\sinh {}^2(\chi
)(d\vartheta
^2+\sin {}^2\vartheta d\varphi ^2)\right)
\end{equation}
from (\ref{Fried}) we obtain
\begin{equation}
\label{state}\frac{d\epsilon }{p+\epsilon }=-3\frac{da}a,
\end{equation}
\begin{equation}
\label{state2}\quad \kappa \epsilon =\frac 3{a^2}\left[ (\frac{da}{dt}%
)^2-1\right]
\end{equation}
where $\kappa $ is the constant appearing in the Einstein's Equations,
$%
\epsilon $ is the energy, and $p$ is the pressure. Assuming that there
is
not interaction between radiation and dust, we have $\epsilon =\epsilon
_d+\epsilon _r$ where, \cite{Chernin}
\begin{equation}
\epsilon _d=\frac A{a^3},\quad \epsilon _r=\frac B{a^4},
\end{equation}
after introducing the conformal time $\eta $ related to $t$ by
\begin{equation}
\label{time}\eta =\int \frac{dt}a
\end{equation}
and integrating $a(\eta ),$ and imposing as initial condition $a(\eta
)\rightarrow b$ when $\eta \rightarrow -\infty ,$ being $b$ a constant,
we
obtain the line element

\begin{equation}
\label{metric}ds^2~=~b^2(e^\eta ~+~1)^2%
\pmatrix{-d\eta ^{2}~+~d\chi ^{2}~+\sinh ^{2}\chi (d\vartheta
^{2}~+~\sin ^{2}\vartheta &d\varphi ^{2})}
\end{equation}

\noindent Among the advantages of the line metric (\ref{metric}) we can
mention that, this line element is regular for any value of the time
parameter $\eta .$ . The homogeneous and isotropic space-time
(\ref{metric})
evolves in permanent expansion towards and asymptotically flat region at
$%
\eta \rightarrow +\infty .$ This asymptotic behavior resembles the one
obtained in the perfect fluid cosmological model \cite{Costa}. In
addition, $%
\partial /\partial \eta $ approaches a timelike Killing vector when
$\eta
\rightarrow -\infty $. Also, we have that, after separating variables in
the
Klein Gordon and Dirac equations, the equations governing the time
dependence are solvable in terms of special functions. It is the purpose
of
the present paper to obtain the quantum spectral distribution of scalar
and
spin 1/2 particles created in the cosmological model associated with the
line element (\ref{metric}) The paper is organized as follows: In \S 2
we
solve the Klein-Gordon equation in the background field (\ref{metric})
and,
in \S 3 we also solve the Dirac equation in this metric. Finally, in \S
4,
based on the results obtained in \S 2 and \S 3, we compute, via the
Bogoliubov coefficients, the probability distribution of the number of
created particles \cite{Parker3}

\section{Solution of the Klein-Gordon equation}

\setcounter{equation}{0}

The covariant generalization of the Klein-Gordon equation reads:

\begin{equation}
\label{2.1}%
\pmatrix{g^{\alpha \beta }\nabla _{\alpha }\nabla _{\beta }-\xi
R-m^{2}}\Phi
{}~=~0
\end{equation}
\noindent where $R$ is the scalar curvature

\begin{equation}
\label{R}R~=~\frac 6{b^2(e^\eta +1)^3}
\end{equation}

\noindent and $\xi $ is a dimensionless parameter which for a
conformally
coupling takes the value of $\xi ~=~1/6.$

\noindent Introducing a solution of the form

\begin{equation}
\label{2.3}\Phi ~=~T(\eta )X(\chi)Y(\vartheta ,\varphi )
\end{equation}
\noindent and substituting (\ref{2.3}) into, (\ref{2.1}) we find that,
after
separating variables, Eq. (\ref{2.1}) reduces to the following set of
ordinary differential equations

\begin{equation}
\left( \frac{d^2}{d\vartheta ^2}+\cot \vartheta \frac d{d\vartheta
}+\frac
1{\sin {}^2\vartheta }\frac{d^2}{d\varphi ^2}\right) Y~=~l(l+1)Y
\end{equation}

\begin{equation}
\label{ecua}\left( \frac{d^2}{d\eta ^2}+2\frac{\dot a}a\frac d{d\eta
}+\frac{%
\ddot a}a+m^2a^2+\sigma ^2-1\right) T=0
\end{equation}
\begin{equation}
\left( \frac{d^2}{d\chi ^2}+2\coth \chi \frac d{d\chi
}+\frac{l(l+1)}{\sinh
{}^2\chi }+\sigma ^2\right) X=0
\end{equation}
where $l$ and $\sigma $ are constants of separation, with $l\geq 0$ and
$%
\sigma ^2>1.$

After introducing the auxiliary function $Z(\eta )$,
\begin{equation}
T(\eta )~=~{\frac{Z(\eta )}{a(\eta )}}\qquad
\end{equation}

Eq. (\ref{ecua}) takes the form:

\begin{equation}
\label{second}\left( {\frac{d^2}{d\eta ^2}}+m^2a^2(\eta )+\sigma
^2-1\right)
Z~=~0
\end{equation}
\noindent substituting the expansion factor $a(\eta )~=~b(e^\eta ~+~1)$
into
(\ref{second}), we obtain the equation,

$$
\left( {\frac{d^2}{d\eta ^2}}+m^2b^2e^{2\eta }~+2m^2b^2e^\eta
+b^2m^2+\sigma
^2-1\right) Z~=~0
$$
\noindent whose solution can be expressed in terms of confluent
hypergeometric functions $M(a,b,z)$ and $U(a,b,z)$ \cite{Abramowitz} as
follows:

$$
Z(\eta )~=~e^{\eta c}\exp (-imbe^\eta )(c_1M({\frac
12}~+~c~+ibm,2c+1,2ibme^%
\eta )
$$
\begin{equation}
+~c_2U({\frac 12}~+~c~+ibm,2c+1,\ 2ibme^\eta )
\end{equation}
\noindent where $c_1$ and $c_2$ are arbitrary constants and $c$ is given
by
the expression

\begin{equation}
\label{c}c~=~-i(\sigma ^2~+~m^2b^2~-~1)^{1/2}
\end{equation}
The solutions of (2.4) are given in terms of the harmonic polynomials:
$%
Y=Y_{lm}(\vartheta ,\varphi )$, also we have that the solution of the
equation (2.6) reads

\begin{equation}
X~=~{\frac 1{\sqrt{\sinh \ \chi }}}P_{i\nu -1/2}^{-l-1/2}(\cosh \chi )
\end{equation}
\noindent where $P_\beta ^\alpha (\cosh x)$ are the torus functions
\cite
{Gradshtein}. and $\nu ^2~=~\sigma ^2-1$.

\section{Solution of the Dirac equation}

\setcounter{equation}{0}

The covariant generalization of the Dirac equation in curved space-time
reads \cite{Brill}

\begin{equation}
\label{3.1}\left\{ \tilde \gamma ^\mu (\partial _\mu -\Gamma _\mu
)+m\right\} \Psi ~=~0
\end{equation}
\noindent where $\tilde \gamma ^\mu $ are the curved gamma matrices
satisfying the relation,

\begin{equation}
\left\{ \widetilde{\gamma }^\mu ,\widetilde{\gamma}^\nu \right\}
_{+}=2g^{\mu \nu }
\end{equation}
\noindent and $\Gamma _\mu $ are the spin connections \cite{Brill}.
Choosing
to work in the diagonal tetrad gauge for $\tilde \gamma ^\mu $,

\begin{equation}
\tilde \gamma ^0=~a^{-1}\gamma ^0,\qquad \tilde \gamma ^1=a^{-1}\gamma
^1,\tilde \gamma ^2~=a^{-1}(\sinh \chi )^{-1}\gamma ^2,
\end{equation}
\begin{equation}
\tilde \gamma ^3~=~a^{-1}(\sinh \chi )^{-1}(\sin \vartheta )^{-1}\gamma
^3
\end{equation}
\noindent where $\gamma ^\mu $ are the standard Dirac flat matrices, we
have
that the Dirac equation (3.1) takes the form

\begin{equation}
\label{Dirac}\left\{ {\frac 1a}\gamma ^0\partial _\eta +{\frac 1a}\gamma
^1\partial _\chi +\frac 1{a\sinh \chi }\gamma ^2\partial _\vartheta
+\frac
1{a\sinh \chi \sin \vartheta }\gamma ^3\partial _\varphi +m\right\} \Psi
=0
\end{equation}
\noindent where we have introduced the spinor $\Psi $,

\begin{equation}
\label{3.5}\tilde \Psi =a^{-2}(\sinh \chi )^{-1}(\sin \vartheta
)^{-1/2}\Psi
\end{equation}
\noindent Applying the algebraic method of separation of variables \cite
{Andrush,Shishkin,Victor} it is possible to write eq. (\ref{Dirac}) as a
sum
of two first order differential operators ${\bf \hat K}_1$ $\ {\bf \hat
K}_2$
satisfying the relation

\begin{equation}
[{\bf \hat K}_1,{\bf \hat K}_2]_{-}=0,\ \left\{ {\bf \hat K}_1+{\bf \hat
K}%
_2\right\} \Phi =0
\end{equation}
\begin{equation}
\label{3.7}{\bf \hat K}_1\Phi =\lambda \Phi =-{\bf \hat K}_2\Phi
\end{equation}
\noindent where

\begin{equation}
\label{3.8}\Psi =\gamma ^1\gamma ^2\gamma ^3\Phi ,
\end{equation}
\begin{equation}
\label{3.9}{\bf \hat K}_1=\pmatrix{\gamma ^{0}\partial _{\eta }+ am
}\gamma
^1\gamma ^2\gamma ^3,
\end{equation}
\begin{equation}
\label{bf}{\bf \hat K}_2=(\gamma ^1\partial _\chi +{\frac 1{\sinh \chi
}}%
\gamma ^2\partial _\vartheta +{\frac 1{\sinh \chi \sin \vartheta
}}\gamma
^3\partial _\varphi )\gamma ^1\gamma ^2\gamma ^3\ {}
\end{equation}
in this way, we have separated the time variable $\eta $ from the
spatial $%
\chi ,\vartheta ,$ and $\varphi $ variables. The problem arises when we
try
to reduce (\ref{bf}$)$ to the form (\ref{3.7}). It is not difficult to
see
that there are not separating matrices allowing that step. In order to
go
further we write equation $({\bf \hat K}_2~+~\lambda )\Phi ~=~0$ as
follows:

\begin{equation}
\label{3.11} \pmatrix{\hat{\bf K}_{3}+\hat{\bf K}_{4}\gamma ^{1}\gamma
^{2}}%
\Phi ~=~0
\end{equation}
\noindent where ${\bf \hat K}_3$ and ${\bf \hat K}_4$ are two commuting
differential operators given by the expressions

\begin{equation}
{\bf \hat K}_3~=~(\sinh \chi )%
\pmatrix{\gamma ^{2}\gamma ^{3}\partial _{\chi }+\lambda },\quad {\bf
\hat K}%
_4~=~\left( \gamma ^2\gamma ^3\partial _\vartheta +\frac 1{\sin
\vartheta
}\partial _\varphi \right) ,
\end{equation}
After introducing the auxiliary spinor $\Sigma $

\begin{equation}
\label{3.13}\Phi ~=~%
\pmatrix{\hat{\bf K}_{4}~+~\gamma ^{1}\gamma ^{2}\hat{\bf K}_{3}}\Sigma
\end{equation}
\noindent we reduce equation (\ref{3.11}) to the form;

\begin{equation}
\left[ {\bf \hat N}_1+{\bf \hat N}_2\right] \Sigma ~=~0,\quad \left[
{\bf %
\hat N}_1,\ {\bf \hat N}_2\right] _{-}~=~0
\end{equation}
\noindent with

\begin{equation}
{\bf \hat N}_2\Sigma =-{\bf \hat N}_1\Sigma =k^2\Sigma
\end{equation}
\noindent where

\begin{equation}
{\bf \hat N}_1~=~(\sinh \chi )^2%
\pmatrix{\partial ^{2}_{\chi }+\coth \chi \partial _{\chi }-\lambda
\gamma ^{2}\gamma ^{3}\coth \chi +\lambda ^{2}}
\end{equation}
\begin{equation}
\label{ene}{\bf \hat N}_2~=~%
\pmatrix{\partial ^{2}_{\vartheta }+i\gamma ^{2}\gamma ^{3}k_{\varphi
}(\sin \vartheta )^{-2}\cos \vartheta -k^{2}_{\varphi }(\sin \vartheta
)^{-2}}
\end{equation}
\noindent $k_\varphi $ in (\ref{ene}) is the eigenvalue of the operator
$%
-i\partial _\varphi $.

\noindent Choosing to work in the following representation of the Dirac
matrices:
$$
\gamma ^0=\left(
\begin{array}{cc}
-i & 0 \\
0 & i
\end{array}
\right) ,\quad \gamma ^1=\left(
\begin{array}{cc}
0 & \sigma ^3 \\
\sigma ^3 & 0
\end{array}
\right) ,
$$

$\quad $%
\begin{equation}
\gamma ^2=\left(
\begin{array}{cc}
0 & \sigma ^2 \\
\sigma ^2 & 0
\end{array}
\right) ,\quad \gamma ^3=\left(
\begin{array}{cc}
0 & \sigma ^1 \\
\sigma ^1 & 0
\end{array}
\right) ,\quad
\end{equation}

\noindent we have that the spinor $\Sigma $ takes the form

\begin{equation}
\label{3.19}\Sigma ~=~\left(
\begin{array}{c}
p(\chi )A(\vartheta ) \\
q(\chi )B(\vartheta ) \\
r(\chi )C(\vartheta ) \\
s(\chi )D(\vartheta )
\end{array}
\right)
\end{equation}
\noindent the functions $p(\chi ),q(\chi ),r(\chi )$ and $s(\chi )$
satisfy
the differential equation

\begin{equation}
\pmatrix{\sinh \chi (d_{\chi }+\mp i\lambda )}%
\pmatrix{\sinh \chi (d_{\chi }+\pm i\lambda )}X(\chi )+k^2X(\chi )~=~0
\end{equation}
\noindent where the upper signs correspond to $p(\chi ),r(\chi )$, and
the
lower signs are related to $q(\chi )$ and $s(\chi )$.

\noindent The functions $A(\vartheta ), B(\vartheta ), C(\vartheta ),
D(\vartheta )$ are solution of the equation

\begin{equation}
\pmatrix{d_{\vartheta }\pm k_{\varphi }(\sin \vartheta )^{-1}}%
\pmatrix{d_{\vartheta }\mp k_{\varphi }(\sin \vartheta
)^{-1}}Y(\vartheta
)-k^2Y(\vartheta )~=~0
\end{equation}
\noindent where the upper signs correspond to $A(\vartheta )$ and $%
C(\vartheta )$, and the lower ones correspond to $B(\vartheta )$ and $%
D(\vartheta )$. After substituting the expression (\ref{3.19}) into
(\ref
{3.13}) we arrive at

\begin{equation}
\label{3.22}\Phi ~=~ik%
\pmatrix{p(\chi )B(\vartheta )(1+i)\alpha \cr q(\chi )A(\vartheta
)(1-i)\alpha \cr r(\chi )D(\vartheta )(1+i)\beta \cr s(\chi )C(\vartheta
)(1-i)\beta\cr }%
e^{ik_\varphi \varphi }
\end{equation}
\noindent where $\alpha $ and $\beta $ are functions depending on the
variable $\eta $. The functions $A,B,C,D$, and $p,q,r,s$ and are related
by
the following systems of coupled partial equations:

\begin{equation}
\label{3.23}(d_\vartheta -{\frac{k_\varphi }{\sin \vartheta }})%
\pmatrix{A\cr C}~=~-k\left(
\begin{array}{c}
B \\
D
\end{array}
\right) ,\quad (d_\vartheta +{\frac{k_\varphi }{\sin \vartheta }})%
\pmatrix{B\cr D}~=~k\pmatrix{A\cr C}
\end{equation}
\begin{equation}
\label{3.24} \sinh \chi (d_\chi +i\lambda )\pmatrix{p\cr r}~=~k%
\pmatrix{q\cr s},\quad \sinh \chi (d_\chi -i\lambda )\pmatrix{q\cr
s}~=-k%
\pmatrix{p\cr r}
\end{equation}
\noindent where $\epsilon $ and $\omega $ are constants. From
(\ref{3.22})
and (\ref{3.9}) we can put $p(\chi )~=~r(\chi )$, $q(\chi )~=~s(\chi )$,
$%
A(\vartheta )~=~C(\vartheta )$, and $B(\vartheta )~=~D(\vartheta )$. The
functions $\alpha $ and $\beta $ satisfy the system:

\begin{equation}
\label{3.25} (\partial _\eta +ima)\beta ~=~-\lambda \alpha ,\qquad
(\partial
_\eta -ima)\alpha ~=~\lambda ~\beta
\end{equation}
\noindent from which we obtain:

\begin{equation}
(d_\eta ^2~+~m^2b^2e^{2\eta }~+e^\eta ~(2m^2b^2-imb)~+~m^2b^2~+~\lambda
^2)\alpha ~=~0
\end{equation}
\begin{equation}
(d_\eta ^2~+~m^2b^2e^{2\eta }~+e^\eta ~(2m^2b^2+imb)~+~m^2b^2~+~\lambda
^2)\beta ~=~0
\end{equation}
\noindent whose solution can be written in terms of confluent
hypergeometric
functions as follows

$$
\alpha (\eta )~=~e^{\eta c}e^{-imbe^\eta }(-c_0{\frac{(c+imb)}\lambda }%
M(c+imb+1,\ 2c+1,2imbe^\eta )
$$
\begin{equation}
+~c_1{\frac{(c^2~+~m^2b^2)}\lambda }U(c+imb+1,2c+1,\ 2imbe^\eta ))
\end{equation}
\begin{equation}
\beta (\eta )~=e^{\eta c}e^{-imbe^\eta }\ \left(
c_0M(c+imb,2c+1,2imbe^\eta
)~+~c_1U(c+imb,2c+1,2imbe^\eta )\right)
\end{equation}

\noindent where $c$ is given by the expression:

\begin{equation}
\label{3.30}c~=~-i(\lambda ^2~+~m^2b^2)^{1/2}
\end{equation}
\noindent The solution of the coupled system of equations (\ref{3.23})
can
be expressed in terms of Jacobi polynomials as follows:

\begin{equation}
A(\vartheta )~=~\sin {}^k\varphi (\vartheta )\cos (\vartheta
/2)P_n^{(k}\varphi ^{+1/2,}{}^k\varphi ^{-1/2)}(\cos \vartheta )
\end{equation}
\begin{equation}
B(\vartheta )~=~\sin {}^k\varphi (\vartheta )\sin (\vartheta
/2)P_n^{(k}\varphi ^{-1/2,}{}^k\varphi ^{+1/2)}(\cos \vartheta )
\end{equation}
\noindent where $n$ is given by the expression

\begin{equation}
n~=~\mid k\mid -\mid k_\varphi \mid -{\frac 12}
\end{equation}
\noindent The system (\ref{3.24}) can be rewritten in the following
matrix
form

\begin{equation}
\left( {\bf I\ }d_\chi +i\lambda \sigma ^3-\frac{i{\it k}}{\sinh \chi }%
\sigma ^2\right) \left(
\begin{array}{c}
p \\
q
\end{array}
\right)
\end{equation}
\noindent after applying the matrix transformation $S$

\begin{equation}
S^{-1}\pmatrix{p\cr q}~=~\pmatrix{P\cr Q},S~=\frac 1{\sqrt{2}}%
\pmatrix{1+i\sigma ^{1}}
\end{equation}
\noindent we reduce (\ref{3.24}) to the form

\begin{equation}
\label{3.36}(d_\chi -{\frac{ik}{\sinh \chi }})Q~=~\lambda P\qquad
(d_\chi +{%
\ \frac{ik}{\sinh \chi }})P~=~-\lambda Q
\end{equation}
\noindent with,

\begin{equation}
P+iQ=~p,\quad Q~+iP~=~q
\end{equation}
\noindent the solution of the system (\ref{3.36}) can be expressed in
terms
of Gauss hypergeometric functions \cite{Gradshtein},

$$
P(\chi )~=~\sinh {}^{ik}(\chi )\cosh (\chi /2)\ {\tt x}
$$
\begin{equation}
F({\frac 12}~+i\lambda ~+ik,{\frac 12}~-i\lambda ~+ik,{\frac
32}+ik,(1-\cosh
\chi )/2)
\end{equation}
$$
Q(\chi )~=~-\pmatrix{({1\over 2}+ik)/\lambda }\sinh {}^{ik}(\chi )\sinh
(\chi /2)\ {\tt x}
$$
\begin{equation}
F({\frac 12}~+i\lambda ~+ik,{\frac 12}~-i\lambda ~+ik,{\frac
12}+ik,(1-\cosh
\chi )/2)
\end{equation}
\noindent Finally, from (\ref{3.5}), (\ref{3.8}) and (\ref{3.22}) we
find
that the solution of the Dirac equation (\ref{3.1}) reads

\begin{equation}
\label{spinor}\tilde \Psi =a^{-2}(\sinh \chi )^{-1}(\sin \vartheta
)^{-1/2}\left(
\begin{array}{c}
p(\chi )B(\vartheta )(i+1)\beta (\eta ) \\
q(\chi )A(\vartheta )(1-i)\beta (\eta ) \\
p(\chi )B(\vartheta )(1+i)\alpha (\eta ) \\
q(\chi )A(\vartheta )(1-i)\alpha (\eta )
\end{array}
\right) e^{ik_\varphi \varphi }
\end{equation}
\noindent Here some comments are in order. The election of a diagonal
tetrad
for separating variables and solving the Dirac equation in the
background
field (\ref{metric}) was done for the sake of simplicity. If we want to
reobtain an angular dependence of the spinor (\ref{spinor}) in terms of
spherical harmonics, it is enough to achieve the transformation relating
the
diagonal tetrad gauge with the Cartesian one \cite{villalba} Obviously,
this
transformation does not affect the time dependence of the spinor
solution.

\section{Particle creation}

\setcounter{equation}{0}

In this section we proceed to analyze the phenomena of scalar and Dirac
particle creation in the cosmological background associated with the
metric (%
\ref{metric}). The model in question begins at $\eta ~=-~\infty $ and
evolves to $\eta \rightarrow +\infty $. Among the particularities of our
model we have to mention that, like \cite{Costa}, the universe
(\ref{metric}%
) is asymptotically flat, but does not present a period of contraction
during its evolution toward the future.

In order to establish the asymptotic behavior of the solutions, at $\eta
{}~=-\infty $ and at $\eta ~=+\infty $, of the corresponding wave
equations,
we write the Hamilton-Jacobi equation

\begin{equation}
g^{\alpha \beta }S,_\alpha S,_\beta ~+~m^2~=~0
\end{equation}
\noindent in the background field (\ref{metric}). Then we obtain:

\begin{equation}
\label{HJ}-(S,_\eta )^2~+~(S,_\chi )^2~+~{\frac 1{\sinh {}^2\chi
}}\left(
(S,_\vartheta )^2~+~{\frac 1{\sin {}^2}}_\vartheta (S,_\varphi
)^2\right)
+~m^2a^2~=~0
\end{equation}
We can separate in Eq. (\ref{HJ}) the time dependence from the spatial
one
as follows:

\begin{equation}
S~=~{\sf T}(\eta )~+~F(\chi ,\vartheta ,\varphi )
\end{equation}
\noindent then, we obtain

\begin{equation}
({\sf T},_\eta )^2~=~m^2a^2~+~\Lambda ^2,\ \rightarrow {\sf T}~=~\pm
\int
(m^2a^2~+~\Lambda ^2)^{1/2}d\eta
\end{equation}
\noindent where $\Lambda $ is a constant of separation. Then, the
quasiclassical asymptotes of the solutions are

\begin{equation}
\label{asy}\Psi \rightarrow F(\chi ,\vartheta ,\varphi )\exp
\pmatrix{\pm i\int (m^{2}a^{2}+\Lambda ^{2})^{1/2}d\eta }
\end{equation}
The asymptotic behavior of the function $T(\eta )$ is

\begin{equation}
\lim _{\eta \rightarrow -\infty }{\sf T}~\rightarrow ~\pm
(m^2b^2+\Lambda
^2)^{1/2}\eta ,\quad \lim _{\eta \rightarrow \infty }{\sf T}\rightarrow
\pm
mbe^\eta ~
\end{equation}
The relation between the conformal time $\eta $ and the cosmic time $t$
can
be obtained from (\ref{time}) giving as result,
\begin{equation}
\label{relation}b(e^\eta +\eta )=t
\end{equation}
notice that the functions $b(e^\eta +\eta )$ monotonically increases,
and $%
t\rightarrow \pm \infty $ when $\eta \rightarrow \pm \infty ,$ and, when
$%
\eta $~$\rightarrow $~+$\infty $  the metric (\ref{metric}) becomes a
Milne
universe \cite{Birrel}

We have that the solutions of the Klein-Gordon equation with the
asymptotic
behavior (\ref{asy})can be written as,

\begin{equation}
\label{4.7}Z_{-\infty }(\eta )~=~(2c)^{-1/2}e^{\eta c}\exp (-imbe^\eta
)M({%
\frac 12}~+~c~-ibm,2c+1,2ibme^\eta )
\end{equation}
\noindent for $\eta $~$\rightarrow $~$-\infty $, and

\begin{equation}
\label{4.8}Z_\infty (\eta )~=~e^{i\pi c/2}e^{\eta c}\exp (-imbe^\eta
)U({%
\frac 12}~+~c~-ibm,2c+1,2ibme^\eta )
\end{equation}
\noindent for $\eta $~$\rightarrow $~$\infty $. where, in order to
analyze
the form (\ref{4.7}) takes at infinity, we have used the formula: \cite
{Abramowitz}

\begin{equation}
\label{4.9}U(\alpha ,\ \gamma ,\ z)\rightarrow z^{-\alpha }\left(
1+O(\frac
1z)\right) ,
\end{equation}
and the relation (\ref{relation})

\noindent Notice that, for the constant $c$ given by (\ref{c}),
$T_{-\infty
} $ and $T_\infty $ can be associated with a positive frequency mode at
$%
\eta ~\rightarrow ~-\infty $ and $\eta ~\rightarrow ~\infty $
respectively.
The next step is to compute the Bogoliubov coefficients relating the two
different vacua at $\eta ~=~-\infty $ and $\eta ~\rightarrow ~\infty $
\cite
{Birrel}. In the present case it is possible to obtain $\beta $ without
performing the integration. Indeed, using the recurrence formula between
the
confluent hypergeometric functions $U(\alpha ,\gamma ,z)$ and $M(\alpha
,\gamma ,z)\ $\cite{Abramowitz},

\begin{equation}
\label{4.10}M(\alpha ,\gamma ,z)~=~{\frac{\Gamma (\gamma )}{\Gamma
(\gamma
-\alpha )}}e^{i\pi \alpha }U(\alpha ,\gamma ,z)~+~{\frac{\Gamma (\gamma
)}{%
\Gamma (\alpha )}}e^{i\pi (\alpha -\gamma )}e^zU(\gamma -\alpha ,\gamma
,-z)
\end{equation}
\noindent substituting $\alpha ~=~{\frac 12}~+~c$~$+ibm$, $\gamma
{}~=~2c+1$
and\ \ $z~=~2ibme^\eta \ $ into (\ref{4.9}) and taking into account the
relation

\begin{equation}
e^zU(\gamma -\alpha ,\gamma ,-z)~=~e^zz^{1-\gamma }(-1)^{1-\gamma
}U(1-\alpha ,2-\gamma ,-z)
\end{equation}
\noindent we arrive at,

$$
Z_{-\infty }(\eta )~=~{\frac{\Gamma (2c+1)}{(2c)^{1/2}\Gamma ({\frac
12}%
{}~+~c~-ibm)}}e^{i\pi /2}e^{i\pi c/2}e^{-\pi mb}Z_\infty (\eta )~+~
$$
\begin{equation}
\label{4.12}+~{\frac{\Gamma (2c+1)}{{(2c)^{1/2}}\Gamma ({\frac
12}~+~c~-ibm)}%
}e^{-i\pi /2}e^{-i\pi c/2}e^{-\pi mb}(2mb)^{-2c}Z_\infty ^{*}(\eta )
\end{equation}
\noindent then, using the relations for the $\Gamma $ function, \cite
{Gradshtein}

\begin{equation}
\label{4.13}2^{2z-1}\Gamma (z)\Gamma (z~+~{\frac 12})~=\sqrt{\pi }\Gamma
(2z)
\end{equation}

\begin{equation}
\mid \Gamma (iy)\mid ^2=\frac \pi {y\sinh (\pi y)}
\end{equation}

\noindent and considering the limit case when $b$ is small, we arrive at

\begin{equation}
\mid \beta \mid ^2~=~{\frac 1{e^{2\pi \nu }-1}}
\end{equation}
\noindent where is $\nu ~=~(\sigma ^2~-~1)^{1/2}$

\noindent Analogously, we have that the Bogoliubov coefficients for the
Dirac case can be obtained by means of the functions $a(\eta )$ and
$\beta
(\eta )$ determining the time dependence of the spinor (\ref{spinor}).
Then,
considering the corresponding positive frequency solutions at the origin
and
at infinity,

\begin{equation}
\beta _{-\infty }(\eta )~=~e^{\eta c}e^{-imbe^\eta
}M(c+imb,2c+1,2imbe^\eta
)
\end{equation}
\begin{equation}
\beta _\infty (\eta )~=~e^{i\pi c/2}e^{\eta c}e^{-imbe^\eta
}U(c+imb,2c+1,2imbe^\eta )
\end{equation}

\noindent where the constant $c$ is given by (\ref{3.30}). Taking into
account the relation (\ref{4.9}) and (\ref{4.10}) we obtain

$$
\beta _0(\eta )~=~{\frac{\Gamma (2c+1)}{\Gamma (c+1+imb)}}e^{i\pi
(c+imb)}e^{-i\pi c/2}\beta _\infty (\eta )~+
$$
\begin{equation}
+~{\frac{\Gamma (2c+1)}{\Gamma (c-imb)}}e^{-i\pi (c+imb+1)}e^{i\pi
c/2}(2mb)^{-2c}\beta _\infty ^{-}(\eta )~
\end{equation}
\noindent then, for small values of $b$, using the relation for the
gamma
function $\Gamma (z),$ \cite{Gradshtein}

\begin{equation}
\mid \Gamma (iy+\frac 12){}\mid ^2={{{\frac \pi {\cosh (\pi y)}}{}{}{}}\
}
\end{equation}

\noindent and the expression (\ref{4.13}) we arrive at:

\begin{equation}
\label{Fermi}\mid \beta \mid ^2~=~{\frac 12}{\frac 1{e^{2\pi \lambda
}+1}}
\end{equation}
\noindent where $\lambda $ \ in (\ref{Fermi}) is the separation constant
appearing in (\ref{3.7})

\section{Concluding remarks}

In this paper we have analyzed the process of creation of scalar and
Dirac
particles in an isotropic and homogeneous cosmological universe filled
by
non interacting radiation and dust-like matter. It is worth noticing
that,
also in this model we have obtained that the average number of particles
created satisfies a thermal distribution law. This result show that the
equation of state does not determine the quantum spectrum but the time
behavior of the line element associated with the cosmological universe.
{}From
a mathematical point of view, this paper shows the capabilities of the
algebraic method of separation of variables and how it works when first
order separation is not possible. We hope that the results presented in
this
article will be of help in understanding quantum phenomena in
cosmological
backgrounds. 


\vspace{1.0 cm}


\begin{thebibliography}{99}
\bibitem{Parker1}  Parker L. 1968 Phys. Rev. Lett. {\bf 21}, 562

\bibitem{Grib}  Grib A. A., Mamaev G. G., and Mostepanenko. V. M. (1970)
Soviet J. Nuclear. Phys. {\bf 10}, 722.

\bibitem{Parker2}  Parker L. 1989 Phys. Rev. Lett. {\bf 59}, 1369

\bibitem{Sahni}  Sahni V. 1988 Mod. Phys. Lett. {\bf 3}, 1425

\bibitem{Birrel}  Birrel N. B. and Davies P. C. W. 1982 {\it Quantum
Fields
in Curved Space} (Cambridge: Cambridge University Press)

\bibitem{Grib2}  Grib A. A., Mamaev G. G., and Mostepanenko. V. M.
Quantum
Vacuum effects in strong fields (Energoatomizdat, Moscow, 1988)

\bibitem{Costa}  Costa I. Deruelle N. Novello M. Svaiter N. 1989 Class.
Quantum. Grav. {\bf 6} 1893

\bibitem{Chernin}  Chernin A. D. 1966 {\it Sov. Astron.} {\bf 9}, 871

\bibitem{Kharbediya}  Kharbediya L. I. 1977 {\it Sov}. {\it Astron}.
{\bf 21}%
, 672

\bibitem{Parker3}  Parker L. 1975 Phys. Rev. D. {\bf 12} 1519

\bibitem{Abramowitz}  Abramowitz M. and Stegun I. A. {\it Handbook of
Mathematical functions} (Dover, New York, 1964)

\bibitem{Gradshtein}  Gradshtein I. S. Ryzhik I. M. Table of integrals,
Series and Products (Academic Press, New York 1980)

\bibitem{Brill}  Brill D and Wheeler J A 1957 Rev. Mod. Phys. {\bf 29}
465.

\bibitem{Andrush}  Andrushkevich I. E and Shishkin G. V 1987 Theor.
Math.
Phys. {\bf 70} 204

\bibitem{Shishkin}  Shishkin G V and Villalba V M 1989 J. Math. Phys.
{\bf 30%
} 2132

\bibitem{Victor}  Shishkin G V and Villalba V M 1992 J. Math. Phys. {\bf
33}
2093

\bibitem{villalba}  Villalba V. M. and Percoco U. 1990 J. Math. Phys.
{\bf 31%
} 715
\end{thebibliography}
\end{document}